\documentclass[11pt]{article}
\usepackage{epsfig}

\newcommand{\be}{\begin{equation}}
\newcommand{\ee}{\end{equation}}

\newcommand{\mw}{M_W}
\newcommand{\mh}{M_H}
\newcommand{\mt}{M_t}
\newcommand{\seff}{s^2_{eff}}
\newcommand{\gl}{\Gamma_\ell}
\newcommand{\equ}[1]{Eq.~(\ref{#1})}

\newcommand{\efe}[1]{Ref.\cite{#1}}

\newcommand{\lsim}{\;\rlap{\lower 3.5 pt \hbox{$\mathchar \sim$}} \raise 1pt
 \hbox {$<$}\;}
\newcommand{\gsim}{\;\rlap{\lower 3.5 pt \hbox{$\mathchar \sim$}} \raise 1pt
 \hbox {$>$}\;}

\def\pr#1#2#3{{\em Phys. Rev. }{\bf D#1~}(19#2)~#3}
\begin{document}
\begin{titlepage}
\begin{flushright}
        \small
\end{flushright}

\begin{center}
\vspace{1cm}
\renewcommand{\thefootnote}{\fnsymbol{footnote}}
{\large\bf Constraining the Higgs boson mass through the combination of
    direct search and precision measurement results\footnote{Contributed paper
  to the Workshop on ``Confidence Limit'', CERN, Geneva, Switzerland, 
  Jan.\ 17-18~2000.}}

\vspace{0.5cm}
{\bf    G.~D'Agostini$^{a,b}$ and  G.~Degrassi$^c$ }
\setcounter{footnote}{0}
\vspace{.8cm}

{\it
   $^a$ Dipartimento di Fisica, Universit{\`a} di Roma ``La Sapienza'', \\
     Sezione INFN di Roma 1, P.le A.~Moro 2, I-00189 Rome,    Italy\\
\vspace{2mm}
   $^b$ CERN, Geneva, Switzerland\\
\vspace{2mm}

        $^c$ Dipartimento di Fisica, Universit{\`a}
                  di Padova, Sezione INFN di Padova,\\ 
                  Via F.~Marzolo 8 , I-35131 Padua, Italy\\
\vspace{1cm} }

{\large\bf Abstract}
\end{center}
\vspace{.2cm} 
We show that the likelihood ratio of Higgs search experiments 
is a form to report the experimental  results suitable to be 
combined with the information from precision measurements 
to obtain a joint constraint on the Higgs mass. We update our
previous combined analysis using the new results on direct searches and
recent precision  measurements, including also the Z$^\circ$ 
leptonic partial width
result. The method is also improved to  take into account small non linearity 
effects in the theoretical formulae. We find an expected value for
the Higgs mass around 160-170 GeV with an expectation uncertainty, quantified 
by the standard deviation of the distribution, of about 50-60 GeV. 
The 95\% probability upper limit comes out to be around 260-290 GeV.
\noindent

\end{titlepage}
%
\setcounter{page}{2}

\section{Introduction}
The Higgs boson is still the missing particle of the Standard Model (SM)
picture and a considerable effort has been devoted to search for evidence of 
it. Unfortunately, till now all direct search experiments have been 
unsuccessful. However, the impressive  amount of data collected at LEP, SLC, 
and the Tevatron allows to probe the quantum structure of the SM, thereby 
providing indirect information about the Higgs mass. While the negative 
outcome of the Higgs searches at LEP 
is usually reported as a combined $95\%$ Confidence Level (C.L.) lower bound, 
the virtual Higgs effects are analyzed through a $\chi^2$  fit to 
the various precision observables that allows a $95\%$ C.L.\ upper bound to be
derived. In \efe{dd} we proposed a method to combine  the information on the
Higgs boson coming from  direct searches (that we indicate generically as 
{\it dir.}) with that obtained from precision measurements ({\it ind.}), 
in order to derive a probability density function (p.d.f.) for  
its  mass
$$
f(m_H\,| \mbox{``data'',``SM''} ) \equiv f(m_H\,|\,\mbox{\it dir.}\, 
 \& \, \mbox{\it ind.}) 
$$
conditioned by both kind of experimental results under the assumption of 
validity of the SM. The heart of our method is the use of the likelihood
of the Higgs search experiments normalized to its value in the
case of pure background, the so called likelihood ratio ${\cal R}$, 
to further constraint the p.d.f. for the Higgs mass obtained employing only 
the precision physics data, $f(m_H\,|\,\mbox{\it ind.})$.

In this paper we would like to recall the main features of 
our method and present an updated analysis. With respect to \efe{dd} we 
improve our analysis in several aspects: i) we use the exact ${\cal R}$ 
for searches up to $\sqrt{s} =196$ GeV as provided by the LEP Higgs Working 
Group \cite{Mac}. ii) We include as observables most sensitive
to the Higgs boson mass
not only the effective mixing parameter, $\sin^2{\theta^{lept}_{eff}}
\equiv \seff$, and the $W$ boson mass, $\mw$, but also 
the $Z^\circ$ leptonic partial width, $\gl$. iii) We take 
into account small non linearity effects in the theoretical formulae. iv) We 
use the most recent results on the various precision observables.

\section{Role of the likelihood ratio in reporting results of searches}
We begin by discussing the role of the likelihood ratio in constraining the
Higgs mass. Let us assume that through the information coming from precision
measurements we have obtained  $f(m_H\,|\,\mbox{\it ind.})$. 
A natural question to ask is then how this p.d.f. should
be modified in order to take into account the knowledge that the Higgs
boson has not been observed at LEP for center of mass energies up to the
highest available. To answer this question let us discuss first an ideal case.
We consider a search for Higgs production in association with a particle of 
negligible width in an experimental  situation of ``infinite'' luminosity,
perfect efficiency and   no background whose outcome was no candidate. 
In this situation we are sure that  all mass values below a  sharp kinematical 
limit $M_K$ are excluded. 
This implies that: a) the p.d.f. for $M_H$  must vanish below $M_K$;
b) above $M_K$ the relative probabilities cannot change, because 
there is no sensitivity in this region, and then the experimental
results cannot give information over there. 
For example, if $M_K$
is $110$ GeV, then $f(200\, \mbox{GeV})/f(120\, \mbox{GeV})$ 
must remain constant before and after 
the new piece of information is included. 
In this ideal case we have then
\begin{equation}
f(m_H\,|\,\mbox{\it dir.} \,\&\,\mbox{\it ind.}) = 
\left\{\begin{array}{ll} 
0   & m_H < M_K  \\
 \frac{f(m_H\,|\,\mbox{\it ind.})}
    {\int_{M_K}^\infty f(m_H\,|\,\mbox{\it ind.})\, \mbox{d}m_H} & m_H \ge M_K\,, 
\end{array}\right.
\label{eq:taglio_secco} 
\end{equation}   
where the integral at denominator is just a normalization coefficient. 

More formally, this result can be obtained making explicit use
of the Bayes' theorem. Applied to our problem, the theorem
can be expressed as follows (apart from a
 normalization constant):
\begin{equation}
f(m_H\,|\, \mbox{\it dir.} \,\&\,\mbox{\it ind.})
 \propto f(\mbox{\it dir.}\,|\,m_H)\cdot  f(m_H\,|\,\mbox{\it ind.})\,,
\label{eq:Bayes}
\end{equation} 
where $f(dir\,|\,m_H)$ is the so called  likelihood. 
In the idealized example  we are considering now, 
$f(dir\,|\,m_H)$ can be expressed in terms of
the probability of  observing zero candidates in an experiment sensitive 
up to a $M_K$ mass for a given value $m_H$, or
\begin{equation}
f(\mbox{\it dir.} \,|\, m_H) = f(\mbox{``zero cand.''}\,|\,m_H) = 
\left\{\begin{array}{ll} 
0   & m_H < M_K \\
      1 & m_H \ge M_K \, . 
\end{array}\right.
\label{eq:lik_taglio_secco} 
\end{equation}   
In fact, we would expect an ``infinite'' number of events 
if $M_H$ were below the kinematical limit.
Therefore the probability of observing nothing should be zero. 
Instead, for $M_H$ above $M_K$,
the condition of vanishing production cross section and no background
can only yield no candidates. 

Consider now a real life situation.  In this case 
the transition between Higgs mass values which are impossible to 
those which are possible is not so sharp. 
In fact because of physical reasons (such as threshold effects
and background) and experimental reasons (such as luminosity
and efficiency) we cannot be  really 
sure about excluding  values close to the kinematical limit, 
nevertheless the ones very far from $M_K$  are ruled out.
Furthermore, the kinematical limit is in general not sharp. In
the case of Higgs production  at LEP  the dominant mode 
is the Bjorken process $ e^+  e^- \rightarrow H + Z^\circ$. 
Indeed, this reaction does not have  a sharp kinematical
limit at $\sqrt{s}-M_Z$ (minus a negligible kinetic energy), due to the 
large total width of the $Z^\circ$. The effective 
kinematical limit ($M_{K_{eff}}$) 
depends on the available integrated luminosity and  could reach up 
to the order of $\approx\sqrt{s}-M_Z+{\cal O}(10\,\mbox{GeV})$ 
for very high luminosity. 
Thus, in a  real life  situation we expect  the ideal step function 
likelihood of  \equ{eq:lik_taglio_secco} to be replaced by a smooth curve 
which goes to zero for low masses. Concerning, instead, the region of no 
experimental sensitivity, $M_H \gsim M_{K_{eff}}$, the likelihood is 
expected
to go to a value  independent on the Higgs mass that however is different
from that of the ideal case, i.e.\ 1, because of the presence of the 
background.

In order to  combine the various pieces of information easily it is  
convenient to replace the likelihood by a function 
that goes to 1 where the experimental sensitivity is 
lost \cite{pia_giulio}. Because constant factors do not play any role in the 
Bayes' theorem this can be achieved by dividing the likelihood by its value
calculated for very large Higgs mass values where no signal 
is expected, i.e. the case of pure background. This likelihood ratio, 
${\cal R}$, can be seen as the counterpart, in the case of a real 
experiment, of the step function of \equ{eq:lik_taglio_secco}. 
Therefore, the Higgs mass p.d.f. that takes into account both 
direct search and precision measurement results can be written as 
\be 
f(m_H\,|\,\mbox{\it dir.} \,\&\,\mbox{\it ind.}) =
\frac{ {\cal R}(m_H)\, f(m_H\,|\,\mbox{\it ind.})}
    {\int_{0}^\infty {\cal R}(m_H)\, f(m_H\,|\,\mbox{\it ind.})\,\mbox{d}m_H}~.
\label{fine}
\ee
In \equ{fine} ${\cal R}$, namely the information from the direct searches,
acts as a shape distortion function of $f(m_H\,|\,\mbox{\it ind.})$. 
As long as ${\cal R}(m_H)$ 
is 1, the shape (and therefore the relative probabilities in that region)
remains unchanged, while ${\cal R}(m_H)\rightarrow 0$ indicates regions 
where the p.d.f. should vanish. 
A {\it conventional} limit can be derived by the ${\cal R}$ function alone
in the transition  region between the region of firm exclusion
(${\cal R}\rightarrow 0$) and the region of insensitivity 
(${\cal R}\rightarrow 1$). However this limit can only have the meaning 
of a `sensitivity bound'~\cite{gda_clw}, and cannot be a 
probabilistic limit which tells us how much we are confident that the Higgs
mass is above a certain value. To express consistently our 
confidence we need to pass necessarely through (\ref{fine}).
 
One should notice that 
${\cal R}(m_H)$ can also assume values larger than 1 for Higgs mass values
below the kinematical limit. This situation corresponds to a number of 
observed candidate events larger than the expected background. In this case 
the role played by
${\cal R}(m_H)$ is to stretch $f(m_H\,|\,\mbox{\it ind.})$ 
below the effective kinematical limit and this might even
prompt a claim for a discovery if ${\cal R}$ becomes sufficiently large 
for the probability of $M_H$ in that region to get 
very close to 1.

\section{Higgs mass inference from precision measurements}
We are going to construct $ f(m_H\,|\,\mbox{\it ind.})$ employing the three 
observables, $\seff$, $\mw$ and $\gl$. These quantities are the most 
sensitive to the Higgs mass and also very accurate measured.  
The most convenient way to approach the problem is to make use 
of the simple parameterization proposed in \efe{Degrassi} and updated in
\efe{DG99} where $\seff,\, \mw$ and $\gl$ are written as functions of
$\mh$, $\mt$, $\alpha_s$ and  the hadronic contribution 
to the running of the electromagnetic coupling:
\begin{eqnarray}
s^2_{eff} &=& (s^2_{eff})_\circ + c_1A_1+c_2A_2-c_3A_3+c_4A_4\,, 
\label{eq:DG3}\\
M_W       &=& M_W^\circ-d_1A_1-d_5A_1^2-d_2A_2+d_3A_3-d_4A_4\, ,
\label{eq:DG4}\\
  \Gamma_\ell & = & \Gamma_\ell^\circ - g_1\, A_1 -g_5\, A_1^2- g_2\, A_2 +
 g_3\, A_3 - g_4\, A_4\, .  
\label{eq:DG5}    
\end{eqnarray} 
In the above equations 
$A_1\equiv \ln(M_H/100\,\mbox{GeV})$,
$A_2\equiv\left[(\Delta\alpha)_h/0.0280-1\right]$, 
$A_3\equiv\left[(M_t/175\,\mbox{GeV})^2-1\right]$
and 
$A_4\equiv\left[(\alpha_s(M_Z)/0.118-1\right]$, where $M_t$ is 
the top quark
mass, $\alpha_s(M_Z)$ is the
strong coupling constant and $(\Delta\alpha)_h$ is the 
five-flavor hadronic
contribution to the QED vacuum polarization at $q^2=M_Z^2$. 
$(s^2_{eff})_\circ$, $M_W^\circ$ and $\gl^\circ$ are (to excellent 
approximation) the theoretical results obtained at the reference point
$(\Delta\alpha)_h=0.0280$, $M_t=175$\,GeV, and $\alpha_s(M_Z)=0.118$ while the
values of the coefficients $c_i$, $d_i$ and $g_i$ are reported in 
Tables 3-5 of \efe{DG99} for three different renormalization scheme.
Formulae (\ref{eq:DG3}--\ref{eq:DG5}) 
are very  accurate for $75 \lsim\mh\lsim 350$ GeV with the
other parameters in the ranges
$170\lsim\mt\lsim 181$ GeV, $0.0273\lsim (\Delta\alpha)_h\lsim
0.0287$, $0.113\lsim \alpha_s(M_Z) \lsim 0.123$. In this case they reproduce 
the exact results of the calculations of 
Refs.(\cite{DGS},\cite{DGV},\cite{DG99}) with maximal errors of 
$\delta s^2_{eff}\sim 1\times 10^{-5}$,
$\delta\mw\lsim 1$ MeV and  $\delta\Gamma_\ell\lsim 3$ KeV, which are all very 
much below the experimental accuracy. Outside the above range, the
deviations increase but remain very small for larger Higgs mass, 
reaching about $3\times 10^{-5}$, 3 MeV, and 4 KeV
at $M_H=600$ GeV for $\seff$, $\mw$, $\Gamma_\ell$, respectively.

Formula (\ref{eq:DG3}) can be seen as providing an indirect measurement
of $A_1= [ \seff - (\seff)_\circ - c_2 A_2 + c_3 A_3 -c_4 A_4]/c_1$,
while (\ref{eq:DG4}) and (\ref{eq:DG5}) of the quantities 
$Y = \mw^\circ - \mw - d_2 A_2 + d_3 A_3 -d_4 A_4$ and
$Z= \gl^\circ - \gl- g_2 A_2 + g_3 A_3 -g_4 A_4 $, 
respectively, all three variables being described by Gaussian p.d.f.'s.
In \efe{dd} we considered only the
relations  (\ref{eq:DG3}) and (\ref{eq:DG4}) and used the fact that the
non linearity effect given by the $d_5$ coefficient is small 
(in the $\overline{MS}$ scheme $d_1 = 5.79\cdot 10^{-2}$, $d_5 = 
8.0\cdot 10^{-3}$) 
to linearize (\ref{eq:DG4}) in $A_1$ in order to directly obtain a $A_1$ 
determination also from $\mw$. Here, instead, we  take into account 
exactly this 
non linearity effect, in view also of the fact that the uncertainty on $\mw$
shrank. 
\begin{figure}
\begin{center}
\begin{tabular}{|c|c|}\hline
 &  \\
$(\Delta \alpha)_h^{EJ}=0.02804(65)$ & 
$(\Delta \alpha)_h^{DH}=0.02770(16)$ \\
&  \\ \hline
\epsfig{file=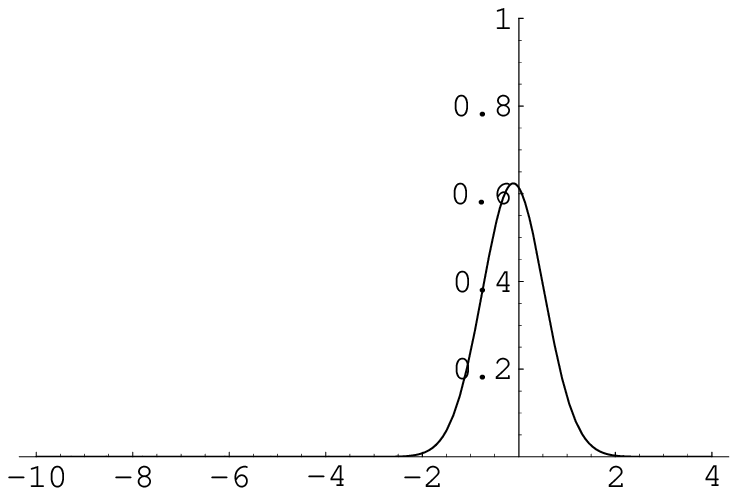,width=0.45\linewidth,clip=} &
\epsfig{file=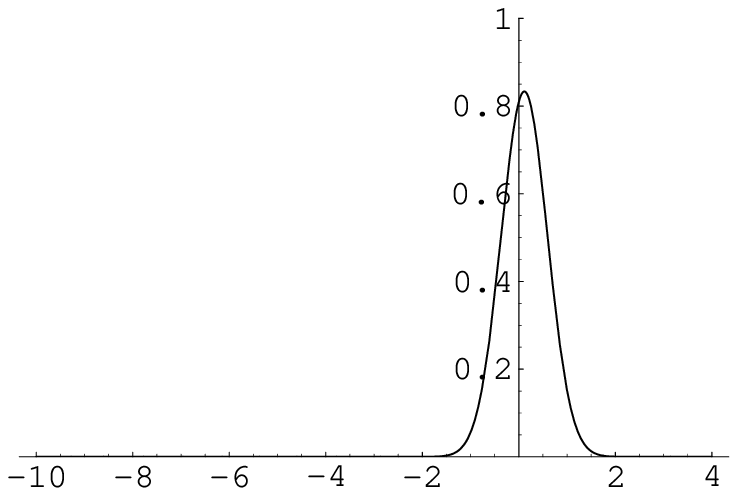,width=0.45\linewidth,clip=} \\ \hline
\epsfig{file=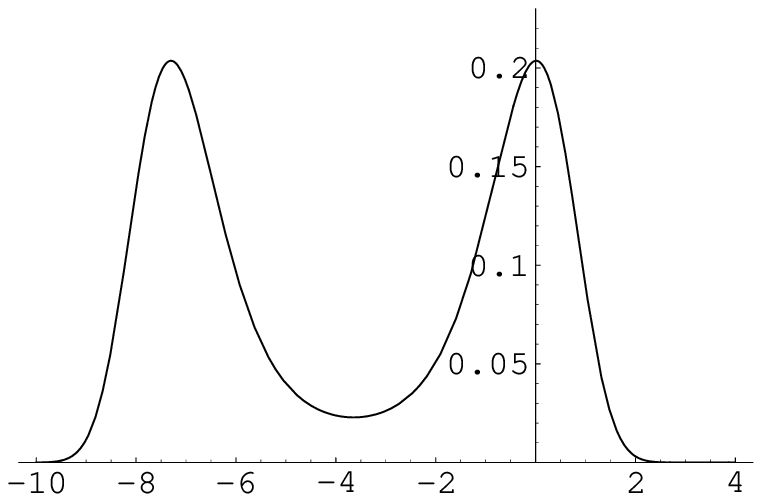,width=0.45\linewidth,clip=} &
\epsfig{file=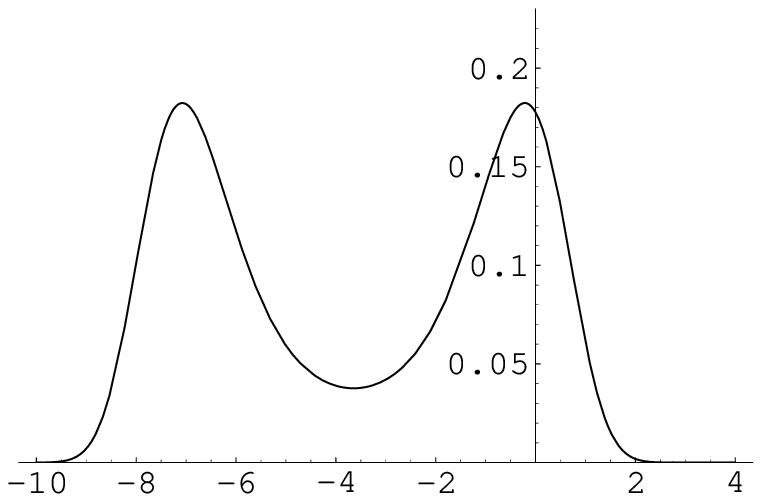,width=0.45\linewidth,clip=} \\ \hline
\epsfig{file=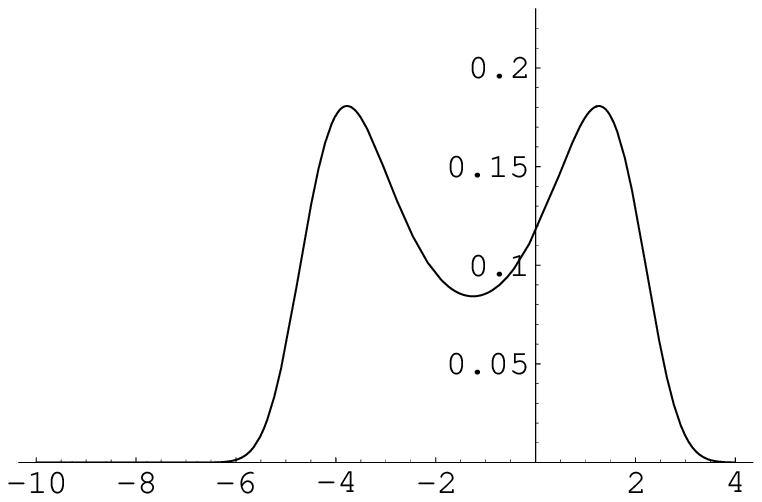,width=0.45\linewidth,clip=} &
\epsfig{file=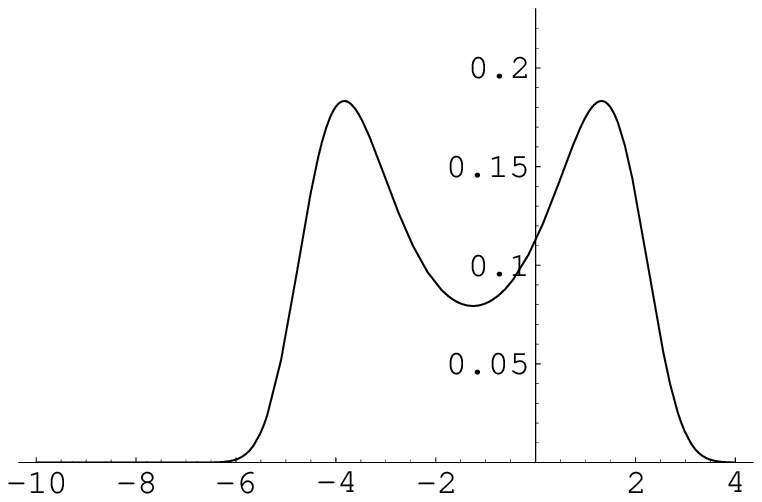,width=0.45\linewidth,clip=} \\ \hline
\epsfig{file=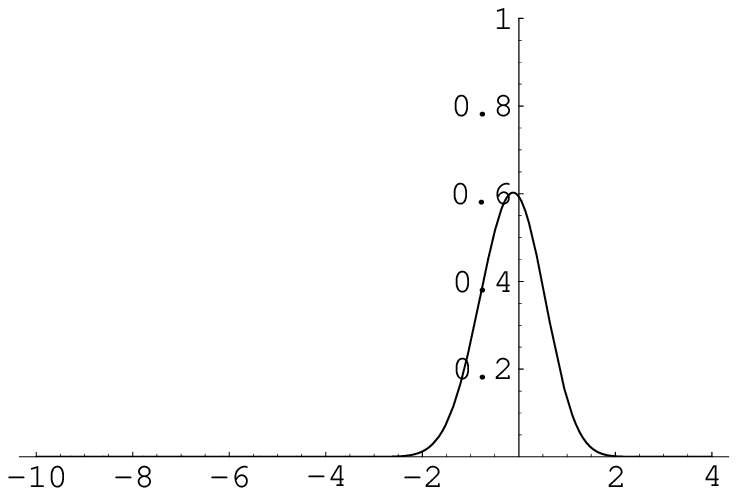,width=0.45\linewidth,clip=} &
\epsfig{file=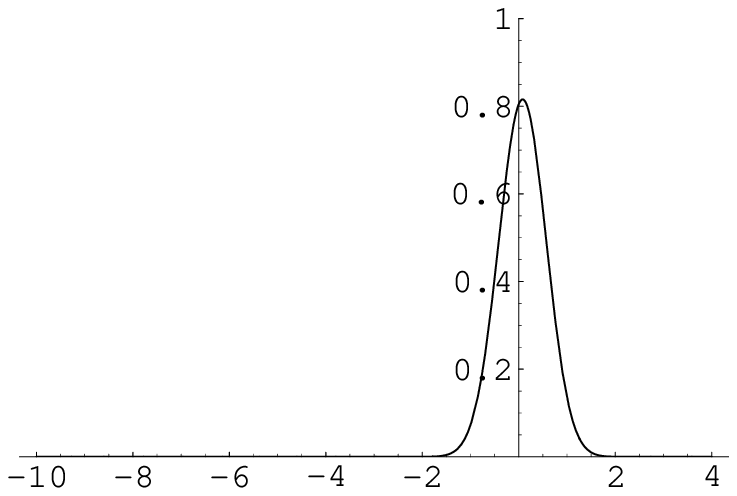,width=0.45\linewidth,clip=} \\ \hline
\end{tabular}
\end{center}
\caption{Probability density function of $A1 \equiv \ln(M_H/100\,\mbox{GeV})$ 
obtained from $s^2_{eff}$, $M_W$, $\Gamma_l$ and their combination
for different values of $(\Delta \alpha)_h$ (see comment on the text 
about low mass peaks).}
\vspace{-17.5cm}\hspace{+5.0cm}$s^2_{eff}$\hspace{+5.5cm}$s^2_{eff}$

\vspace{+3.6cm}\hspace{+5.0cm}$M_W$\hspace{+5.5cm}$M_W$

\vspace{+3.6cm}\hspace{+5.1cm}$\Gamma_l$\hspace{+5.9cm}$\Gamma_l$

\vspace{+3.6cm}\hspace{+5.0cm}All\hspace{+5.8cm}All
\vspace{+7.0cm}
\label{fig:inferences}
\end{figure}
In Fig.~\ref{fig:inferences} we plot the p.d.f.~of $A_1$
 using formulae (\ref{eq:DG3}--\ref{eq:DG5})
and the input quantities specified in the next section. The 
inference from $\seff,\, \mw,$ and $\gl$, separately,   and from their 
combination
is shown. The quadratic expression in (\ref{eq:DG4}) and (\ref{eq:DG5}) for
$\mw$ and $\gl$, respectively, gives rise to an unphysical peak on the
left side of the plot for values of the Higgs mass where  
formulae (\ref{eq:DG3}--\ref{eq:DG5}) are not valid. Neverthless, as
shown in the bottom plot, when the total combination is taken these
unphysical peaks disappear and the inference obtained is concentrated
in the physical region.

The $A_1$, $Y$ and $Z$  determination are clearly correlated, 
therefore one has to built a covariance matrix. This can be easily done
because formulae (\ref{eq:DG3}--\ref{eq:DG5}) are linear in the common
terms $\underline{X}\equiv \{\mt,\,\alpha_s(M_Z),\, (\Delta\alpha)_h\}$. 
The likelihood 
of our indirect measurements $\underline{\Theta} \equiv \{A_1,\, Y, \, Z \}$
is then a three dimensional correlated normal with covariance matrix
\be V_{ij} =\sum_l \frac{\partial \Theta_i}{\partial X_l}\cdot 
\frac{\partial \Theta_j}{\partial X_l} \cdot\sigma^2(X_l)
\label{eq:corrmat}
\ee
or 
\be
f(\underline{\theta} \,|\, \ln(m_H)) \propto e^{- \chi^2/2}
\label{eq: indlik}
\ee
where $\chi^2 = \underline{\Delta}^T {\bf V}^{-1} \underline{\Delta}$ with 
$\underline{\Delta}^T = \{ a_1 - \ln(m_H/100),\,
y - d_1 \ln(m_H/100) -d_5 \ln^2(m_H/100),\,
z - g_1 \ln(m_H/100) -g_5 \ln^2(m_H/100)\}$.
Using Bayes' theorem  the likelihood (\ref{eq: indlik}) can be turned
into a p.d.f. through  the choice 
of a prior. The natural choice is a uniform prior in $\ln(m_H)$, 
since it is well understood that radiative corrections measure
this quantity (for this reason the  likelihood (\ref{eq: indlik})
has been expressed in terms of $\ln(m_H)$). 
Moreover, this choice recovers the result of 
Ref.~\cite{dd}, which was obtained as uncertainty propagation without
explicit use of a prior on the Higgs mass. 
The uniform prior in $\ln(m_H)$ implies 
that $f(\ln(m_H/100) \,|\,\mbox{\it ind.})$ 
is just the normalized likelihood (\ref{eq: indlik}).
Using standard probability calculus we can express our 
results as a p.d.f. of $M_H$:
\be 
f(m_H\,|\,\mbox{\it ind.}) = \frac{ m_H^{-1}\, e^{- (\chi^2/2)}}
         {\int_{0}^\infty m_H^{-1}\, e^{- (\chi^2/2)}\, \mbox{d}m_H}.
\ee

The theoretical coefficients, $c_i,\,d_i,\,g_i$, depends on the renormalization
scheme in which the relevant calculations are done and their numerical
spread is usually taken as an indication of the theory uncertainty of
the calculations. A way to take into account this uncertainty is
to consider different inferences,
each conditioned by a given set of parameters, labelled by $R_i$.
For each renormalization scheme $R_i$ we can construct a
$f(m_H\,|\,\mbox{\it ind.}, R_i)$ and obtain a 
p.d.f.  ``integrated'' over the possible schemes, through
\begin{equation}
f(m_H\,|\,\mbox{\it ind.}) =\sum_i f(m_H\,|\,\mbox{\it ind.}, R_i)
\cdot f(R_i)\,,
\label{eq:fA1av}
\end{equation}
where $f(R_i)$ is the probability assigned to each scheme
($f(R_i)=1/3$ $\forall\, i$). 
The calculation of expectation value and variance
is then straightforward or
\begin{eqnarray}
\mbox{E}[M_H] &=& \frac{1}{3}\sum_i \mbox{E}[M_H\,|\,R_i] \\
\sigma^2(M_H) &=& \frac{1}{3}\sum_i \sigma^2(M_H\,|\,R_i) +
                  \frac{1}{3}\sum_i\mbox{E}^2[M_H\,|\,R_i]-\mbox{E}^2[M_H]\\
              &=& \frac{1}{3}\sum_i \sigma^2(M_H\,|\,R_i) + 
                  \sigma^2_E\,,
\end{eqnarray}
where $\sigma_E$ indicates the standard deviation calculated from the 
dispersion of the expected values. In Ref.~\cite{dd} we have shown 
that results almost identical are obtained if one employs for the coefficients
$c_i,\,d_i$ an average value and a standard deviation
evaluated from the dispersion 
of the values obtained from the various renormalization schemes. 
\section{Results}
The experimental inputs  we use to construct 
$f(m_H\,|\,\mbox{\it ind.})$
are \cite{LEPEWWG}: $\seff=0.23151\pm 0.00017$, $\mw=80.394\pm 0.042$ GeV, 
$\Gamma_\ell=83.96\pm 0.09$ MeV, $\mt=174.3\pm 5.1 $GeV,
$\alpha_s(M_Z)=0.119\pm0.003$. Concerning  $(\Delta\alpha)_h$ in the recent
years there has been a lot of activity on this subject with several 
evaluations. They can be classified as of two types: 
i) the most  
phenomenological analyses,
that rely on the use of  all the available experimental data on the
hadron production in $e^+ \,e^-$ annihilation and on perturbative QCD (pQCD)
for the high energy tail ($E \geq 40$ GeV) of the dispersion integral.
The reference value in this approach is 
$(\Delta\alpha)_h^{EJ} = 0.02804\pm0.00065$~\cite{Jeg}.
ii) The so called ``theory driven'' analyses \cite{jeger,DH}, that
differ from the previous most phenomenological ones
mainly by the use of pQCD down to 
energies of the order of a few GeV and by the treatment of 
old experimental data in regions where pQCD is not applicable. 
The combination of these two factors gives a result  for $(\Delta\alpha)_h$
that differ from  type i) one by
a drastically reduced uncertainty but, at the same time,  a lower
central value. The most stringent evaluation 
of these theory oriented analyses is 
$(\Delta\alpha)_h^{DH}=0.02770\pm0.00016\,$~\cite{DH}, 
that we use   as reference value for this kind of 
approach. 
At the moment there is no definite argument for choosing one or other
of the two approaches.  
The results are absolutely compatible to each other. However,
the numerical difference between  central values and 
uncertainties is such that it prevents  an
easy estimation of the effect 
of choosing one value instead 
of the other.
For these reasons we decided 
to present our results 
for the values of $(\Delta\alpha)_h$ given by 
$(\Delta\alpha)_h^{EJ}$  and  $(\Delta\alpha)_h^{DH}$ 
separately.

The values of the ${\cal R}$ function that enters in \equ{fine} has been
provided by the LEP Higgs Working Group \cite{Mac}: they take into account
the Higgs searches  by all four LEP collaborations for center of mass energy
up to $\sqrt{s} = 196$ GeV. 

\begin{table}[t]
\begin{center}
\begin{tabular}{|l|cc|cc|}\hline
& & & &   \\
  & \multicolumn{2}{|c|}{$(\Delta\alpha)_h= 0.02804(65)$} 
  & \multicolumn{2}{|c|}{$(\Delta\alpha)_h= 0.02770(16)$} \\
& & & &   \\ \hline
& & & &   \\
& $ (ind.)$ 
& $\left(\begin{array}{c} ind. \\ + \\ dir. \end{array}\right)$ 
&$(ind.)$ 
& $\left(\begin{array}{c} ind. \\ + \\ dir. \end{array}\right)$\\
& & & &   \\
$\mbox{\bf E}[\mathbf{M_H}]$/TeV  &  0.10   & {\bf 0.17}  &  0.12 &{\bf 0.16}\\
$\mathbf{\sigma(M_H)}$/TeV        &  0.07   & {\bf 0.06}  &  0.06 &{\bf 0.05}\\
$\hat{M}_H$/TeV                   &  0.06   & 0.11        &  0.09 &0.11\\
$\mathbf{M^{50}_H}$/TeV       &  0.09   & {\bf 0.15}  &  0.11 &{\bf 0.14}\\
& & & &   \\
$P(M_H\le 0.11\,\mbox{TeV})$          &67\,\% & 8.9\,\%   & 51\,\% & 8.5\,\% \\
$P(M_H\le 0.13\,\mbox{TeV})$          &76\,\% & 34\,\%    & 65\,\% & 35\,\%  \\
$P(M_H\le 0.20\,\mbox{TeV})$          &92\,\% & 79\,\%    & 91\,\% & 83\,\%  \\
& & & &   \\
$\mathbf{M^{95}_H}$/TeV; $P(M_H\le M^{95}_H)\approx 0.95$ 
                                          &0.23 &{\bf 0.29}& 0.23& {\bf 0.26}\\
$\mathbf{M^{99}_H}$/TeV; $P(M_H\le M^{99}_H)\approx 0.99$ 
                                          &0.33 &{\bf 0.41}& 0.31& {\bf 0.34}\\
& & & &   \\
$\left\{\!\!\begin{array}{l} M_1/\mbox{TeV};\ P(M_H < M_1)\approx 0.16 \\
                 M_2/\mbox{TeV};\ P(M_H > M_2)\approx 0.16 \end{array}\right.$ 
& $\left\{\!\!\begin{array}{l} 0.05 \\ 0.15 \end{array}\right.$ 
& $\left\{\!\!\begin{array}{l} 0.12 \\ 0.22 \end{array}\right.$ 
& $\left\{\!\!\begin{array}{l} 0.07 \\ 0.17 \end{array}\right.$ 
& $\left\{\!\!\begin{array}{l} 0.12 \\ 0.20 \end{array}\right.$ \\
& & & &  \\
\hline  
\end{tabular}
\end{center}
\caption{\sf Summary of the direct plus indirect information.}
\label{tab:results}
\end{table}

\begin{figure}[t]
\begin{center}
\epsfig{file=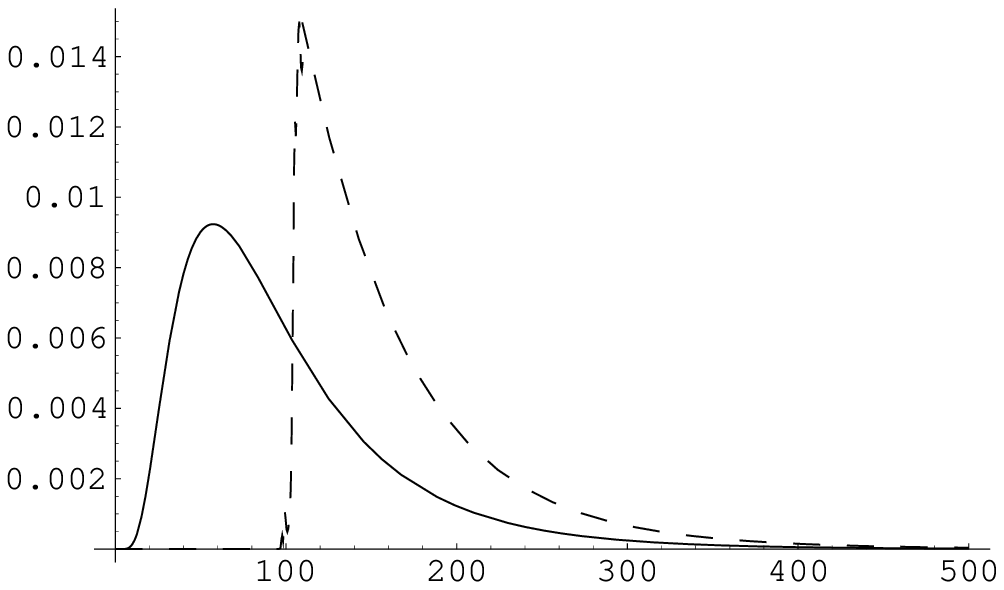,width=0.55\linewidth}\\[0.2cm]
\epsfig{file=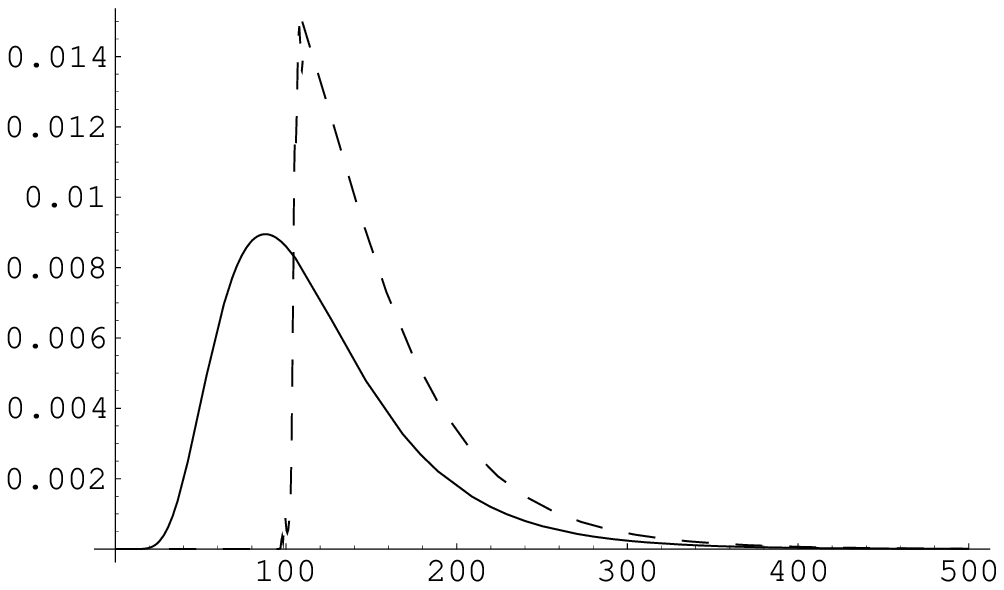,width=0.55\linewidth}
\end{center}
\caption{\sf Probability distribution functions using only indirect information
(solid line) and employing also the experimental results from direct searches
(dashed one): a) $(\Delta \alpha)_h = 0.02804(65)$; b) 
$(\Delta \alpha)_h = 0.02770(16)$.}

\vspace{-10.2cm}\hspace{+3.8cm}\mbox{\scriptsize{$f(M_H)$}}

\vspace{+0.5cm}\hspace{+7.8cm}\mbox{a)}

\vspace{+2.0cm}\hspace{+10.2cm}\mbox{\scriptsize{$M_H$}}

\vspace{+0.4cm}\hspace{+3.8cm}\mbox{\scriptsize{$f(M_H)$}}

\vspace{+0.5cm}\hspace{+7.8cm}\mbox{b)}

\vspace{+1.9cm}\hspace{+10.2cm}\mbox{\scriptsize{$M_H$}}

\vspace{1.9cm}
\label{fig:Jeg}
\end{figure}

Table \ref{tab:results} summarizes the result of our analysis
in terms of various convenient 
parameters of the distribution. We present
the two cases  $(\Delta\alpha)_h=(\Delta\alpha)_h^{EJ}$ and 
$(\Delta\alpha)_h=(\Delta\alpha)_h^{DH}$ and report all values in TeV to 
reduce the number of digits to the significant ones. The shape  of the
p.d.f. with and without the inclusion of the direct search information
is presented in Fig.~\ref{fig:Jeg}.
From this figure one can  notice that, in the case of  
$f(\mbox{\it ind.} \,|\, m_H)$, the use
of a higher central value for $(\Delta\alpha)_h$ 
(i.e.~$(\Delta\alpha)_h^{EJ}$)\ tends to concentrate more the
probability towards smaller values of $M_H$. As a consequence the 
analysis based on $(\Delta\alpha)_h^{EJ}$  gives
results for the standard deviation of the  $M_H$ p.d.f. and 
95\,\% probability upper limit, $M_H^{95}$, very close to those obtained
using $(\Delta\alpha)_h^{DH}$ regardless the fact that the uncertainty
on $(\Delta\alpha)_h^{EJ}$ is 
approximately $4$ times larger than that of  $(\Delta\alpha)_h^{DH}$.
As expected, the inclusion  of the direct search information in the
Higgs mass probability analysis drifts the p.d.f.~towards higher
values of $M_H$, changing its shape such that the probability of
$M_H$ values below 110 GeV drops to $\approx 9 \, \%$.
The various parameters of the distribution (expected value, 
standard deviation, mode ($\hat{M}_H$) and median ($M^{50}_H$))
 are not very sensitive to the values of 
the hadronic contribution to the vacuum polarization. Also in both cases, 
$\approx 80\, \%$ of
the probability is concentrated in the region $M_H < 0.20$ TeV.
Instead the choice of $(\Delta \alpha)_h$ affects the tail of the
distribution with $(\Delta \alpha)_h^{EJ}$ producing a much longer one.
\section{Conclusions}
The likelihood ratio ${\cal R}$ is a form to report the experimental results
that allows an easy  combination of  the information from Higgs 
search experiments  with that coming from precision measurements and accurate 
calculations in order to constraint jointly the Higgs mass. The 
${\cal R}$ function 
is very convenient for comparing and combining the various informations
and it has also an intuitive interpretation because of its limit to 
the step function of the ideal case.  It can be
also seen as the p.d.f. for the Higgs mass in the case of complete lack of 
other information, i.e.\ when one assumes $ f(m_H\,|\,\mbox{\it ind.})=1$.
Although in this case the normalization integral in \equ{fine} is 
mathematically ``infinite'', still the relative probabilities of different 
intervals of mass regions are perfectly well defined. 
However, it is obvious that it is not possible to evaluate  
from the ${\cal R}$ function alone a probabilistic lower limit. 
This can only be done when ${\cal R}$ is combined with  
$f(m_H\,|\,\mbox{\it ind.})$ obtained from
precision measurements, which has the important role of making 
large values of $M_H$ impossible, thus making the final p.d.f. 
normalizable. 

The analysis we have performed clearly shows that a heavy Higgs scenario is
highly disfavored, the data preferring a 
$\log_{10}(M_H/\mbox{GeV})\approx {\cal O}(2)$ . Note that our results
are derived under the assumption of the validity of the SM and rely on
the input experimental and theoretical quantities stated in the text.
All these assumptions seem  to us very reasonable.  
In particular, we don't consider  strong evidence against the SM  the
fact that about one half of $f(m_H\,|\,\mbox{\it ind.})$ is eaten up by
the LEP direct search. 
\vspace{1cm} \\

\noindent We wish to thank the LEP Higgs Working Group for presenting
the likelihood
ratio values of the Higgs searches and P.~Igo-Kemenes and G.~Ganis for useful 
communications.


\begin{thebibliography}{ref99}
\bibitem{dd}
G. D'Agostini and G. Degrassi, Euro.~Phys.~Journal
              {\bf C10}, (1999) 663.

\bibitem{Mac}
P.~McNamara, Report of the LEP Higgs Working Group 
               to LEP Experiments Committee Meeting,
               CERN-Geneva, Sept.~7th 1999.

\bibitem{pia_giulio}
P. Astone and G. D'Agostini, CERN-EP/99-126, {\tt hep-ex/9909047}.

\bibitem{gda_clw}
G. D'Agostini, Contribution to the Workshop on `Confidence Limits', 
CERN, Geneva, 17--18 January 2000.

\bibitem{Degrassi}
G.\ Degrassi, P.\ Gambino, M.\ Passera and A. Sirlin,  Phys. 
               Lett. {\bf B418}, (1998) 209.

\bibitem{DG99} G. Degrassi and, P. Gambino, 
              preprint DFPD-99/TH/19, TUM-HEP-333/98 ({\tt hep-ph/9905472}), 
              to appear on Nucl.~Phys.~{\bf B}.

\bibitem{DGS} G. Degrassi, P. Gambino, and A. Sirlin,  Phys. 
 Lett. {\bf B394} (1997) 188.

\bibitem{DGV} G. Degrassi, P. Gambino, and A. Vicini,  Phys. Lett. 
{\bf B383} (1996) 219.


\bibitem{LEPEWWG} The LEP and SLD Electroweak Working Group, preprint
CERN-EP/99-15, February 1999; for the most recent preliminary results see
{\tt http://www.cern.ch/LEPEWWG}.


\bibitem{Jeg}
S. Eidelman, F. Jegerlehner, Z. Phys. {\bf C67} (1995) 585;
H. Burkhardt, B. Pietrzyk, Phys. Lett. {\bf B356} (1995) 398.

\bibitem{jeger}
A.D.~Martin and D.~Zeppenfeld, Phys.~Lett.~{\bf B345} (1994) 558;
M.~Davier and A.~H\"ocker, Phys.~Lett.~{\bf B419} (1998) 419;
             J.H.~K\"uhn and M.~Steinhauser, Phys.~Lett.~{\bf B437} (1998) 425;
             S.~Groote, J.G.~K\"orner, K.~Schilcher and
             N.F.~Nasrallah, Phys.~Lett.~{\bf B440} (1998) 375;
             J.~Erler, \pr{59}{99}{054008},
             F.\,Jegerlehner, hep-ph/9901386.

\bibitem{DH} 
M.~Davier and A.~H\"ocker, Phys. Lett. {\bf B435} (1998) 427.


\end{thebibliography}
\end{document}